# Are Machine Learning Interatomic Potentials Truly Practical?

## A Benchmark of 23 Mainstream Models

Hanwen Kang[1,2], Tenglong Lu[1,3], Sheng Meng[1]*, Miao Liu[1,3]*

[1] *Institute of Physics, Chinese Academy of Sciences, Beijing 100190, China*

[2] *School of Physical Sciences, University of Chinese Academy of Sciences, Beijing 100049, China*

[3] *Dongguan Institute of Materials Science and Technology, Dongguan, Guangdong 523808, China*

**Corresponding author: smeng@iphy.ac.cn, mliu@iphy.ac.cn*

---

**Abstract.** Most MLIP benchmarks reward static accuracy while ignoring inference efficiency and hardware scalability — driving model bloat with unclear real-world value. We benchmark 23 mainstream open-source MLIPs on a low-cost NVIDIA DGX Spark (128 GB native memory, capped at 80 GB to mimic ordinary lab hardware), using a fixed 192-atom system under a unified ASE-based pipeline. We evaluate three dimensions: predictive accuracy, MD simulation throughput, and atomic scalability. Our results expose a sharp accuracy–efficiency trade-off: large SOTA models deliver only **3–5 meV/atom** more accuracy than lightweight ones, but lose orders of magnitude in throughput — in the worst case, becoming only marginally faster than DFT itself. Lightweight MLIPs, by contrast, sit on the Pareto frontier and run on modest hardware. The lesson is that single-dimensional benchmarks mislead the field, and that future MLIP development should value efficiency and scalability alongside accuracy.

---

## 1. Introduction

Atomistic molecular dynamics (MD) is how materials science has, for half a century, asked what atoms actually do at room temperature. The catch has always been the choice between two inadequate tools.

**Density functional theory (DFT)** [1–3] delivers quantum-level accuracy but scales unfavorably — its cost grows roughly with the **cube** of the number of atoms ($O(N^3)$). A few hundred atoms is comfortable; a few thousand is a research quarter. **Classical empirical force fields** scale linearly and are blazingly fast, but they cannot see the subtle environment-dependent bonding that defines much of materials chemistry. For decades, the field has lived with this trade-off.

Machine-learning interatomic potentials (MLIPs) were supposed to dissolve it. By encoding local atomic environments within a 6–7 Å cutoff and learning high-dimensional potential-energy surfaces from massive DFT datasets, modern MLIPs approach DFT accuracy (energy MAEs of 20–30 meV/atom; force errors in the tens of meV·Å$^{-1}$) while inheriting force-field-like linear scaling [4] [5]. A diverse open-source ecosystem — **Equiformer** [6]**, DPA** [7–9]**, PET** [10,11]**, M3GNet** [12]**, CHGNet** [13]**, MatterSim** [14]**, GRACE** [15,16]**, MACE** [17,18]**, GPTFF** [19]**, MatRIS** [20]**, NequIP** [21–23]**, TACE** [24]**, EquFlash, eSEN** [25]**,SevenNet** [26–28]**, ORB** [29,30]**, Nequix** [31,32]**, Allegro** [33] — has proliferated, each nudging the Matbench Discovery leaderboard a little higher.

And then something went quietly wrong.

While accuracy improved, model size exploded. State-of-the-art universal MLIPs now exceed **730 million parameters** [11] — a scale that dwarfs the size of individual training datasets. The accuracy gains from this expansion are real but small: roughly **3–5 meV/atom**, smaller than room-temperature thermal noise. Meanwhile, inference throughput drops, memory consumption rises, and the maximum size of simulatable systems collapses. In the most extreme cases, large MLIPs run **less than 2x faster than DFT**, surrendering the central efficiency advantage that justified their existence.

This paper is a systematic benchmark study designed to expose that trade-off. We test 23 mainstream open-source MLIPs on standardized hardware and ask three questions: how accurate are they really? how fast? and how big a system can they actually simulate? Our goal is not to crown a winner — it is to provide a clear, multi-dimensional view of the current MLIP landscape, and to give working researchers a practical basis for choosing models.

---

## 2. Methods

All benchmarks were performed on an **NVIDIA DGX Spark** desktop platform with 128 GB of native GPU memory. To reflect the hardware available to an ordinary research group, we capped memory-scaling tests at **80 GB**. Cross-hardware validation was performed on a standard **NVIDIA A100 (80 GB)**. DFT reference calculations used a dual-socket Intel Xeon Gold 6530 server with 64 physical CPU cores and 256 GB system memory.

All simulations ran through the **Atomic Simulation Environment (ASE)** [34] under a unified configuration. To rule out framework bias, we cross-validated a representative subset on **TorchSim** [35], which delivered only a 1.11–1.4x speedup over ASE — confirming that ASE is sufficient for reliable large-scale benchmarking.

We evaluated each model along three dimensions. The first was **accuracy**, quantified by comparing predicted thermal-conductivity properties against DFT ground truth under the official Matbench Discovery protocol. Rather than relying solely on energy MAE, we used the phonon thermal-conductivity metric κ_SRME from Matbench Discovery [36] as the primary accuracy coordinate.

$$\kappa_{SRME} = SRME(\{K_{qs}\}) = \frac{2}{N_c\, V} \cdot \frac{\sum_{q,s} \left| K_{qs}^{\mathrm{MLIP}} - K_{qs}^{\mathrm{DFT}} \right|}{\kappa^{\mathrm{MLIP}} + \kappa^{\mathrm{DFT}}}$$

Energy-based errors depend strongly on the reference DFT functional and data-generation protocol; values obtained under different reference settings, such as PBE [37] and r2SCAN [38], are not directly interchangeable. Because lattice thermal conductivity is determined by harmonic and anharmonic phonon responses, this metric probes the smoothness and higher-order derivative quality of the learned potential-energy surface, not only the local energy value.

The second was **throughput**, measured as the number of valid MD integration steps per unit time and normalized to reflect real trajectory-generation capacity. The third was **hardware scalability**, characterized as the maximum simulatable system size under the unified 80 GB memory cap and extrapolated from a linear memory-vs-atom fit.

The 23 models tested cover the full parameter spectrum. The **lightweight tier** (0.5–5 M parameters) features minimal per-atom memory, very high throughput, and the ability to handle ultra-large systems. The **medium tier** (5–10 M parameters) sits between the two extremes, offering balanced accuracy and cost for routine MD. The **large tier** (10–730 M parameters) is pre-trained on extensive OAM / OMat24 datasets [39,40], delivers top static accuracy, and carries a heavy computational footprint. All models were tested with default official configurations and public pre-trained weights, to ensure fairness and reproducibility.

---

# 3. Results

## 3.1 Accuracy–Efficiency Pareto Trade-off

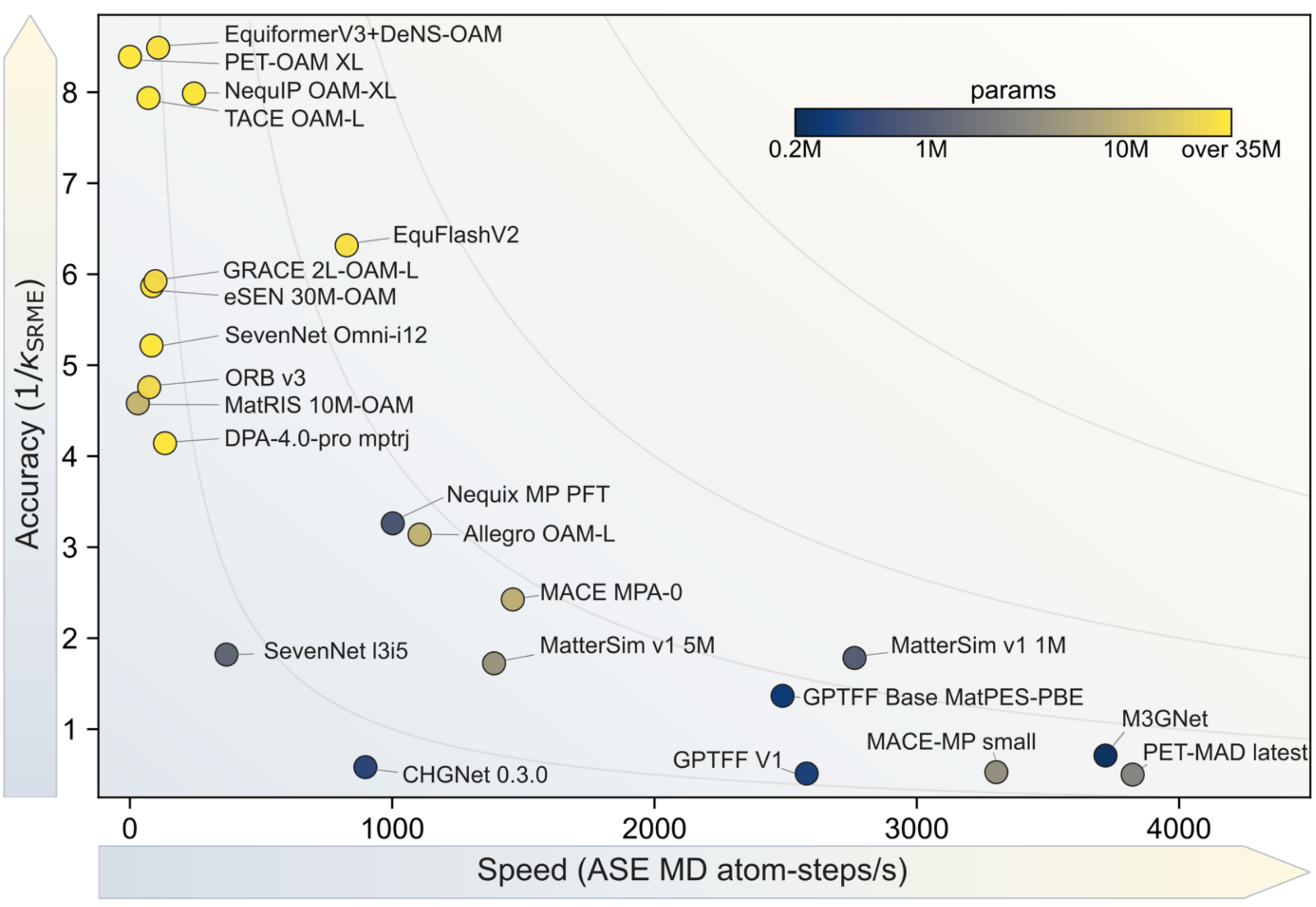

**Figure 1. Accuracy–efficiency Pareto trade-off across the 23 benchmarked MLIPs. Inference throughput (x-axis, normalized MD steps per second) is plotted against prediction accuracy (y-axis, normalized thermal-conductivity MAE). Accuracy and computational efficiency constitute a trade-off.**

The headline result is shown in Figure 1 — a Pareto plot of throughput against prediction accuracy across all 23 architectures. The picture is unambiguous: **the field's biggest, most accurate models are also the slowest.** Across the entire benchmark, parameter scale and inference throughput are tightly and inversely correlated. There is no region of the plot in which a heavyweight model offers both the highest throughput and the highest accuracy; the two metrics trade off against each other almost everywhere.

**Lightweight models** , including MatterSim v1 1M and Nequix MP PFT, sit firmly on the **Pareto frontier**. Their accuracies are not the highest in the field, but they are competitive with much larger models, and they deliver throughputs that are one to three orders of magnitude greater than anything in the large tier. For the vast majority of practical MD simulations — phase transformations, grain-boundary dynamics, finite-temperature defect studies — they offer more than enough accuracy at a cost a desktop machine can bear [41] [42] [43].

**Medium-scale models** , including MACE-MPA-0 and Allegro-OAM-L, also lie on the Pareto frontier. They buy a small accuracy increment over the lightweight tier, at a modest speed cost. The trade is sometimes worth it for problems where every meV matters.

**Large-scale models** — including Equiformer V3, DPA4.0 Pro, and PET-OAM-XL — lie in the upper-right region of the plot: highest accuracy, lowest throughput. Critically, the accuracy advantage they offer over the lightweight tier is only on the order of **3–5 meV/atom,** below thermal noise. They indeed achieve improved phonon-accuracy scores, with $\kappa_{SRME} \approx 0.12$–$0.24$ compared with $\approx 0.31$–$0.58$ for the strongest sub-5M baselines. Below the zero-point vibrational energy of a typical bond. For any experiment a researcher would actually run, that advantage is **statistically indistinguishable from zero**. They pay for this invisible gain with catastrophic throughput loss — between one and three orders of magnitude slower than the lightweight tier. For most users, that is not a trade worth making.

## 3.2 Hierarchical Simulation Speed

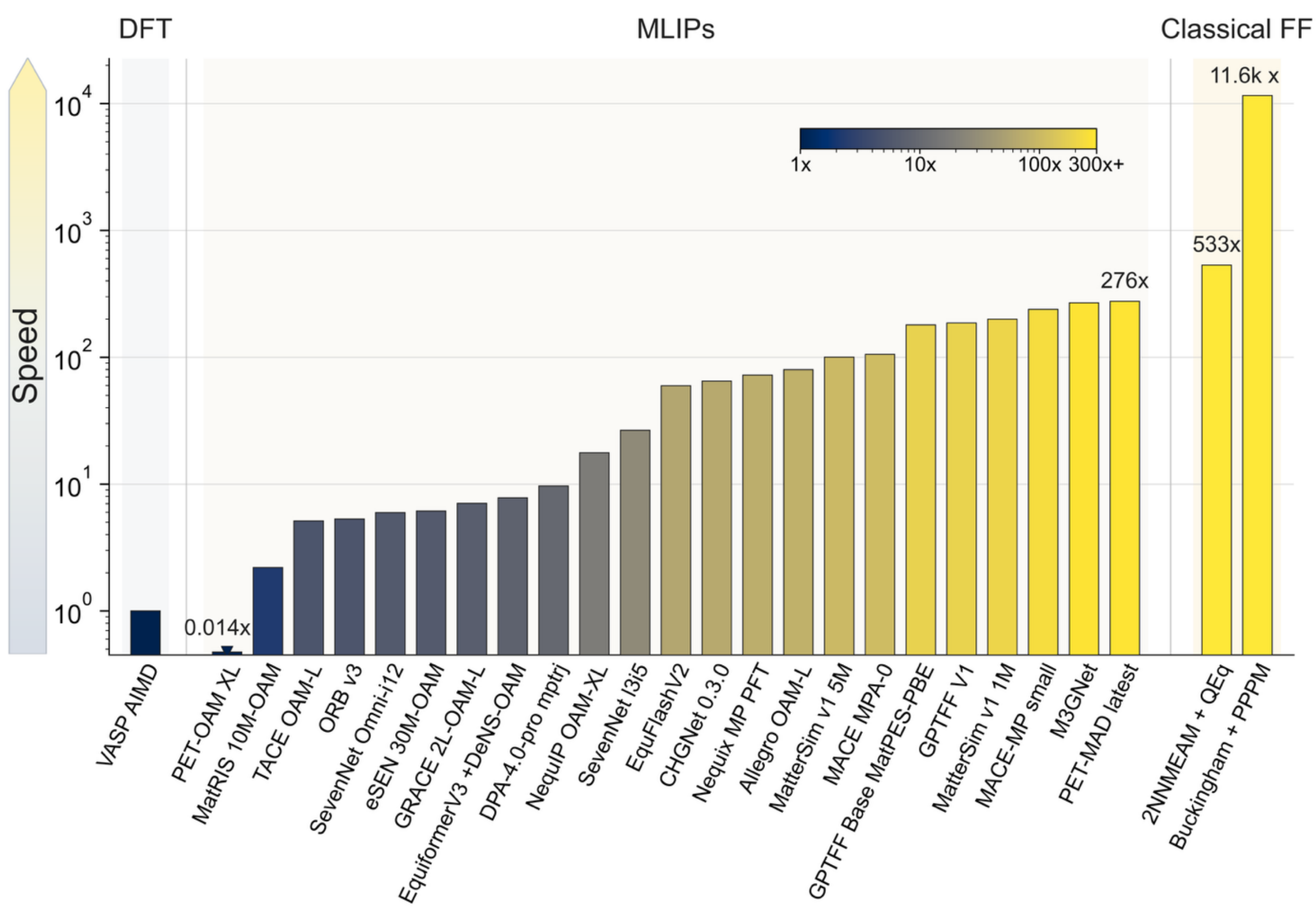

**Figure 2. MD Simulation throughput across multiple mainstream MLIPs, normalized to CPU-based DFT (1×). Classical empirical potentials lead: the Buckingham potential achieves ~11,600× DFT speedup, but at the cost of chemical fidelity. Lightweight MLIPs follow, delivering hundreds-to-thousands-fold speedups — the speed tier at which the original MLIP promise is realized. Medium-scale MLIPs lose roughly half an order of magnitude in throughput. Heavyweight MLIPs collapse to less than 2x DFT in the worst case, surrendering the central efficiency advantage of ML over quantum mechanics.**

Figure 2 ranks simulation paradigms by throughput, with CPU-based DFT as the baseline. The hierarchy is striking and worth dwelling on.

For the speed benchmark, we used a 192-atom LiCoO2 cell computed with GGA-PBE, a 520 eV plane-wave cutoff, and a Gamma-only 1 × 1 × 1 k-point mesh as the common DFT test system. This normalization is useful but should be interpreted with care. DFT cost grows approximately cubically with atom count, whereas most fixed-cutoff MLFFs and classical force fields scale close to linearly. Thus, DFT-normalized speedups provide a practical reference scale, but the more rigorous comparison is between MLFFs and classical force fields under the same atomistic workload.

At the very top sit the **classical empirical potentials**. The Buckingham potential [44], for instance, runs roughly **11,600x faster than DFT**. It is the speed champion of all traditional schemes — but it achieves that speed by sacrificing the chemical fidelity that DFT and MLIPs are designed to provide. It cannot see bond formation, cannot describe pressure-induced phase transitions, and breaks down outside its narrow fitting range. The more complex Li-Co-O 2NNMEAM + QEq potential [45], which performs charge equilibration at every MD step, runs about 533× faster than DFT.

Below the empirical tier sit the **lightweight MLIPs**. They run hundreds to thousands of times faster than DFT — slower than the empirical potentials, but still fast enough to make nanosecond-scale MD of tens of thousands of atoms routine. This is the speed tier at which the original promise of MLIPs is actually delivered: quantum-level accuracy at near-linear cost.

Below that, the picture darkens. **Medium-scale MLIPs** lose perhaps half an order of magnitude in throughput relative to the lightweight tier, while gaining a small accuracy bump. **Heavyweight MLIPs** drop further still. In the worst cases — the very largest architectures, run on the very largest systems they can handle — they run **less than 2x faster than DFT**. The efficiency edge that motivated the entire MLIP program is essentially gone for these models.

The story, in other words, is not that MLIPs are slow. It is that some MLIPs — specifically the largest, most leaderboard-obsessed ones — have abandoned the very reason MLIPs exist.

### 3.3 GPU Memory Scalability

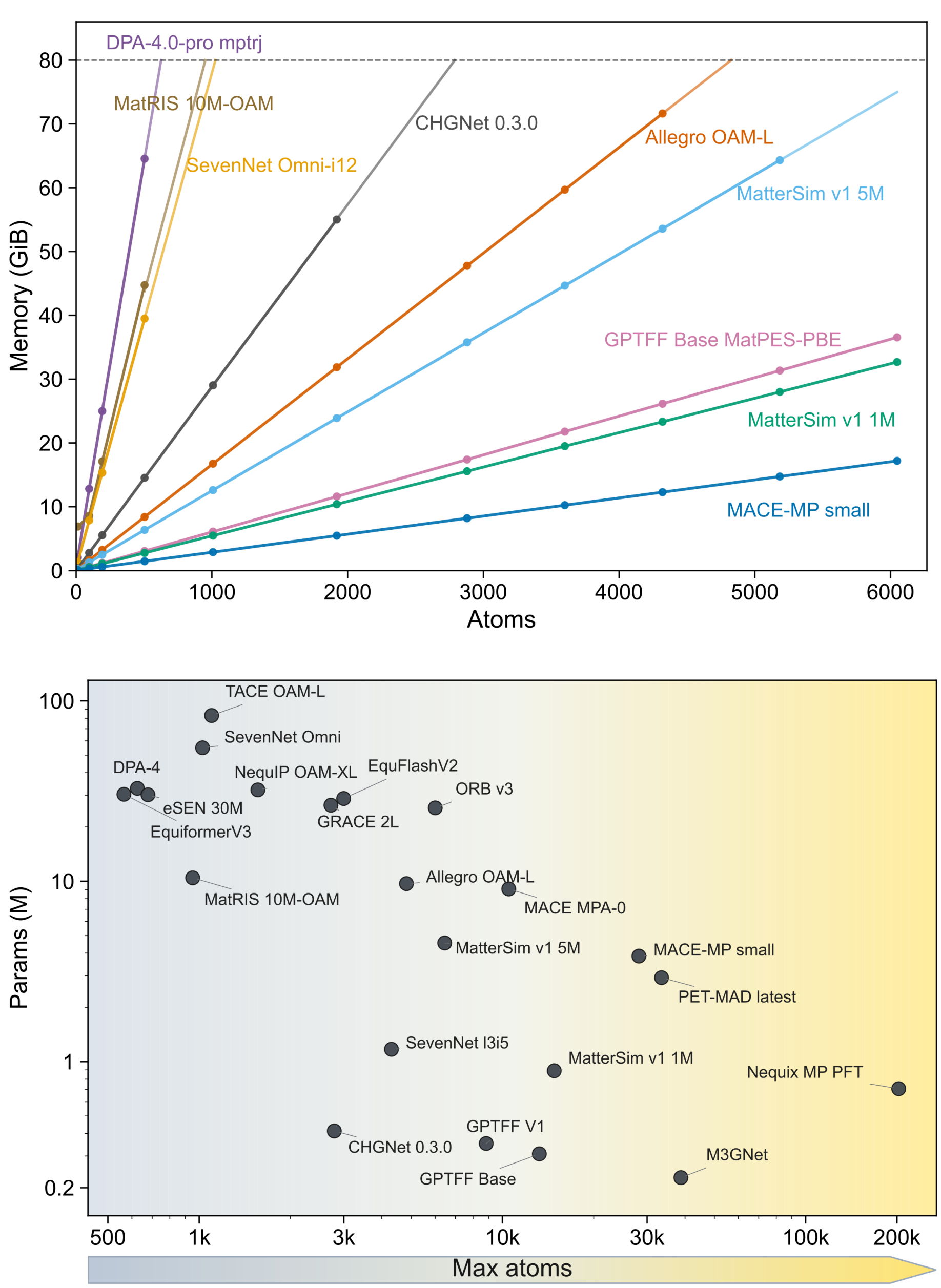

**Figure 3.(a) GPU memory consumption versus simulated atom count, with linear fits extrapolated to the unified 80 GB memory cap.(b) Relationship between maximum atom count and model parameters. Per-atom memory footprints differ by orders of magnitude across architectures. Lightweight MLIPs scale to roughly 200,000 atoms; medium-tier models (e.g., M3GNet) handle $10^4$–$10^5$ atoms; heavyweight models — DPA4.0 Pro, Equiformer V3, eSEN 30M ,PET-OAM XL — saturate at 500–1,000 atoms, comparable to the efficency of DFT calculations.**

Figure 3 plots per-model memory consumption against the number of simulated atoms, and reports the maximum simulatable system size under the 80 GB cap. The plot's central message is that **MLIP memory cost scales linearly with system size, but the slope differs by orders of magnitude across architectures.**

**Lightweight MLIPs** have very low per-atom memory footprints, on the order of a fraction of a megabyte per atom at typical batch sizes. Under the 80 GB cap, they routinely scale to systems of tens of thousands of atoms, and the best of them — Nequix in particular — reach **roughly 200,000 atoms**. That is enough to model a real grain boundary, a real phase transformation, a real material at a meaningful length scale.

**Medium-scale models** sit in the middle. Some, like MACE, are well-optimized and approach the lightweight tier in memory efficiency. Others, however, leave substantial room for optimization: their per-atom memory footprint is higher than it needs to be given their parameter count, suggesting that future work on this tier could pay large dividends.

**Heavyweight models** — and here we name them explicitly: **DPA4.0 Pro, Equiformer V3, and PET-730M** — suffer from severe memory bottlenecks. Under the same 80 GB cap, they are limited to systems of **500 to 1,000 atoms**. That is not a material. It is a small nanoparticle, and not a particularly large one. These models cannot, in any practical sense, simulate the systems that MLIPs were invented to study.

## 3.4 Material Density Matters Too

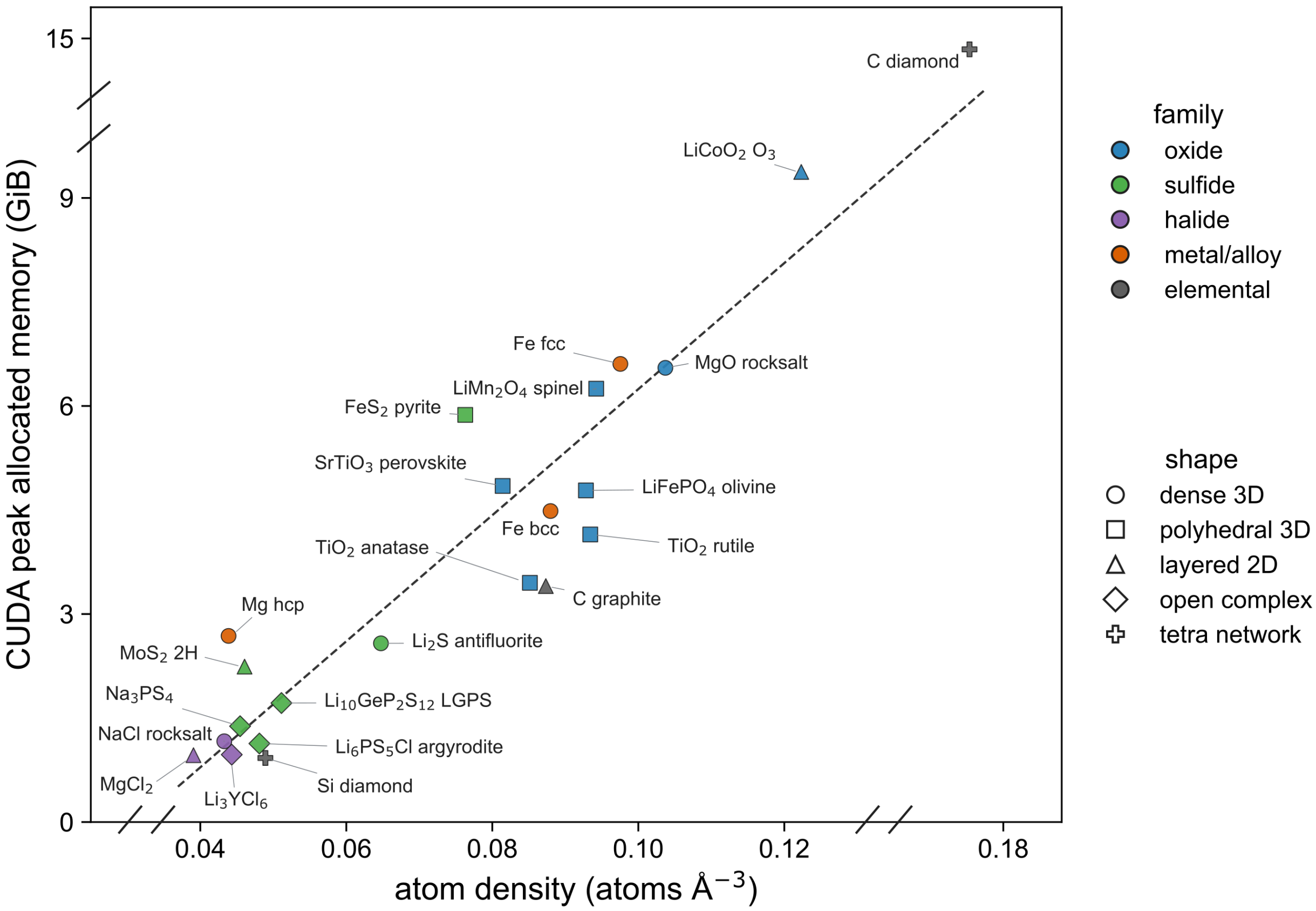

**Figure 4. Material-density-dependent MLIP scalability, comparing dense and sparse test systems on identical hardware. Because GNN-based MLIPs use a fixed 6–7 Å cutoff, dense structures generate expensive graphs while sparse structures generate cheap ones. The same model can simulate on the order of $10^5$ atoms of $Li_{10}GeP_2S_{12}$(LGPS) but only ~8000 atoms of $LiCoO_2$ — two orders of magnitude of headroom locked in the material rather than the model. MLIP scalability is therefore jointly determined by model architecture and target material.**

Figure 4 explores a subtle and underappreciated effect: **the same model on the same hardware scales very differently depending on the material's density.** MLIP scalability is not a property of the model alone — it is a property of the model *meeting* the material.

The mechanism is straightforward. GNN-based MLIPs describe each atom by the contents of a small sphere — a fixed cutoff of 6–7 Å — and ask the network to summarize that sphere into a single embedding vector. The cost of doing so depends almost entirely on **how many neighbors live inside that sphere**, because each neighbor adds edges and nodes to the graph the network must process at every time step.

For example, although $LiCoO_2$ is 2.5 times denser than $Li_{10}GeP_2S_{12}$ (LGPS), GPTFF graph construction amplifies this difference: its neighbor count is 2.94 times larger, and the number of three-body terms is 9.89 times larger. Consequently, under an 80 GB memory budget, GPTFF can simulate on the order of $10^5$ atoms for LGPS, but only ~8,000 atoms for $LiCoO_2$.Two orders of magnitude of headroom, locked away in the structural properties of the material rather than the architecture of the model.

This finding has an immediate practical implication. **Choosing an MLIP without knowing the target material is choosing blind.** Two researchers with the same model and the same GPU may experience radically different scalability depending on whether they simulate a dense oxide or a sparse electrolyte. Benchmark numbers published without specifying the material should be treated with caution.

---

# 4. Discussion

## 4.1 Hardware Validation

The DGX Spark is, in our experience, a credible and inexpensive MLIP benchmarking platform. Although its native memory is 128 GB, we held all tests to 80 GB to match what ordinary laboratories can afford. Cross-validation on an A100 (80 GB) yielded essentially equivalent results, with at most a one-fold speed difference — not an order-of-magnitude gap. For most research groups, a **$4,000-$5,000 desktop unit** is sufficient for serious MLIP MD work. Combined with a standardized ASE / TorchSim pipeline, this makes reproducible, low-cost benchmarking accessible to anyone, and breaks the assumption that meaningful MLIP research requires H100-class hardware.

## 4.2 Why the Field Got Here

The temptation, after reading the results above, is to blame the model developers. They bloated their models, after all. They are the ones shipping hundreds-of-millions-parameter behemoths onto PyPI.

But the models did exactly what the field rewarded them for doing.

**Matbench Discovery** — the de-facto industry leaderboard — measures **one thing**: static prediction accuracy. It does not measure throughput. It does not measure memory. It does not measure whether the model can run a real MD simulation at finite temperature, with a real time step, in a real box. **A benchmark that optimizes one dimension will produce a field that optimizes one dimension.** Four years of single-dimensional optimization have produced hundreds-of-millions-parameter Ferraris that, in fact, cannot leave the parking lot.

The path forward is not to abandon accuracy benchmarks — they have served the field well, and they remain an essential guardrail against underfit models. It is to **complement them with efficiency and scalability metrics**, so that the next generation of models is judged on the problems they actually solve, not just the offline scores they achieve. A multi-dimensional evaluation framework — accuracy, throughput, atomic scalability, density sensitivity — would have caught the present mismatch years ago.

## 4.3 Practical Model Selection

For readers who need a rule of thumb, the recommendations from our benchmarks can be summarized as follows. For large systems, long trajectories, and most practical MD work, the **lightweight tier** — M3GNet, MACE-small, and other models in the 1–5 M parameter range — is the right choice. For mid-sized problems where a little more accuracy matters, the **medium tier** (5–10 M parameters) offers a reasonable middle ground. For researchers who feel compelled to chase the top of the Matbench Discovery leaderboard, the **large tier** (10–730 M parameters) is generally **not recommended**: heavy models such as DPA4-MP trajectory and Equiformer V3 (DNS-OAM) offer essentially invisible accuracy gains while imposing severe computational and memory bottlenecks, and lose the central advantage of MLIPs over DFT. Researchers who adopt them should do so only with a specific need in mind, and with a clear understanding of what they are giving up.

## 4.4 Limitations

This study is not the last word on MLIP benchmarking. All tests ran on a fixed 192-atom system under a unified 80 GB cap, which makes quantitative numbers dependent on system size, material density, and hardware configuration. Future work should expand across multi-scale and multi-material benchmarks, and explore adaptive evaluation criteria for diverse simulation scenarios. Targeted optimization of lightweight MLIP architectures — improving their accuracy without sacrificing efficiency — is, in our view, the most promising direction for the field’s next phase.

---

## 5. Conclusions

We benchmarked 23 mainstream MLIPs across accuracy, throughput, and scalability, on standardized hardware with a conservative memory budget. The headline is unflattering: **largest does not mean best.** Bloated architectures deliver a few meV/atom of extra accuracy while surrendering orders of magnitude in throughput and the ability to simulate real materials. Lightweight MLIPs, sitting comfortably on the Pareto frontier, are the practical choice for nearly all routine MD work.

Beyond a leaderboard, real progress in MLIPs will be measured by a different yardstick: **the ability to ask bigger questions and get honest answers with less effort.** As low-cost, high-memory desktop machines like the DGX Spark become commonplace, that yardstick is well within reach. The next generation of MLIPs will not be remembered for topping Matbench. It will be remembered for running all weekend on a researcher's laptop — and being trusted.

[1] P. Hohenberg and W. Kohn, Inhomogeneous Electron Gas, Phys. Rev. **136**, B864 (1964).

[2] W. Kohn and L. J. Sham, Self-Consistent Equations Including Exchange and Correlation Effects, Phys. Rev. **140**, A1133 (1965).

[3] G. Kresse and J. Hafner, *Ab initio* molecular dynamics for liquid metals, Phys. Rev. B **47**, 558 (1993).

[4] G. Benedini, A. Loew, M. Hellström, S. Botti, and M. A. L. Marques, Universal machine learning potentials for systems with reduced dimensionality, AI Sci. **1**, 025005 (2025).

[5] M. Liu and S. Meng, Recent breakthrough in AI-driven materials science: tech giants introduce groundbreaking models, Mater. Futures **3**, 027501 (2024).

[6] Y.-L. Liao, A. J. Hoffman, S. C. Shen, A. Duval, S. W. Norwood, and T. Smidt, *EquiformerV3: Scaling Efficient, Expressive, and General SE(3)-Equivariant Graph Attention Transformers*, arXiv:2604.09130.

[7] J. Zeng et al., DeePMD-kit v2: A software package for deep potential models, The Journal of Chemical Physics **159**, 054801 (2023).

[8] J. Zeng et al., DeePMD-kit v3: A Multiple-Backend Framework for Machine Learning Potentials, J. Chem. Theory Comput. **21**, 4375 (2025).

[9] H. Wang, L. Zhang, J. Han, and W. E, DeePMD-kit: A deep learning package for many-body potential energy representation and molecular dynamics, Computer Physics Communications **228**, 178 (2018).

[10] S. N. Pozdnyakov and M. Ceriotti, *Smooth, Exact Rotational Symmetrization for Deep Learning on Point Clouds*, arXiv:2305.19302.

[11] F. Bigi, P. Pegolo, A. Mazitov, J. Schmidt, and M. Ceriotti, *Pushing the Limits of Unconstrained Machine-Learned Interatomic Potentials*, arXiv:2601.16195.

[12] C. Chen and S. P. Ong, A universal graph deep learning interatomic potential for the periodic table, Nat Comput Sci **2**, 718 (2022).

[13] B. Deng, P. Zhong, K. Jun, J. Riebesell, K. Han, C. J. Bartel, and G. Ceder, CHGNet as a pretrained universal neural network potential for charge-informed atomistic modelling, Nat Mach Intell **5**, 1031 (2023).

[14] H. Yang et al., *MatterSim: A Deep Learning Atomistic Model Across Elements, Temperatures and Pressures*, arXiv:2405.04967.

[15] Y. Lysogorskiy, A. Bochkarev, and R. Drautz, *Graph Atomic Cluster Expansion for Foundational Machine Learning Interatomic Potentials*, arXiv:2508.17936.

[16] A. Bochkarev, Y. Lysogorskiy, and R. Drautz, Graph Atomic Cluster Expansion for Semilocal Interactions beyond Equivariant Message Passing, Phys. Rev. X **14**, 021036 (2024).

[17] I. Batatia, S. Batzner, D. P. Kovács, A. Musaelian, G. N. C. Simm, R. Drautz, C. Ortner, B. Kozinsky, and G. Csányi, The design space of E(3)-equivariant atom-centred interatomic potentials, Nat Mach Intell **7**, 56 (2025).

[18] I. Batatia, D. P. Kovács, G. N. C. Simm, C. Ortner, and G. Csányi, MACE: Higher Order Equivariant Message Passing Neural Networks for Fast and Accurate Force Fields, (2022).

[19] F. Xie, T. Lu, S. Meng, and M. Liu, GPTFF: A high-accuracy out-of-the-box universal AI force field for arbitrary inorganic materials, Science Bulletin **69**, 3525 (2024).

[20] Y. Zhou, S. Hu, X. Zhang, H. Wang, G. Tan, and W. Jia, MATRIS: TOWARD RELIABLE AND EFFICIENT PRE- TRAINED MACHINE LEARNING INTERATOMIC POTEN-, (2026).

[21] C. W. Tan et al., High-performance training and inference for deep equivariant interatomic potentials, Digital Discovery **5**, 1558 (2026).

[22] S. Batzner, A. Musaelian, L. Sun, M. Geiger, J. P. Mailoa, M. Kornbluth, N. Molinari, T. E. Smidt, and B. Kozinsky, E(3)-equivariant graph neural networks for data-efficient and accurate interatomic potentials, Nat Commun **13**, 2453 (2022).

[23] B. Kozinsky, A. Musaelian, A. Johansson, and S. Batzner, *Scaling the Leading Accuracy of Deep Equivariant Models to Biomolecular Simulations of Realistic Size*, in *Proceedings of the International Conference for High Performance Computing, Networking, Storage and Analysis* (ACM, Denver CO USA, 2023), pp. 1–12.

[24] Z. Xu, W. Xie, and P. Hu, *Spectral/Spatial Tensor Atomic Cluster Expansion with Universal Embeddings in Cartesian Space*, arXiv:2509.14961.

[25] X. Fu, B. M. Wood, L. Barroso-Luque, D. S. Levine, M. Gao, M. Dzamba, and C. L. Zitnick, *Learning Smooth and Expressive Interatomic Potentials for Physical Property Prediction*, arXiv:2502.12147.

[26] J. Kim, J. Kim, J. Kim, J. Lee, Y. Park, Y. Kang, and S. Han, Data-Efficient Multifidelity Training for High-Fidelity Machine Learning Interatomic Potentials, J. Am. Chem. Soc. **147**, 1042 (2025).

[27] Y. Park, J. Kim, S. Hwang, and S. Han, Scalable Parallel Algorithm for Graph Neural Network Interatomic Potentials in Molecular Dynamics Simulations, J. Chem. Theory Comput. **20**, 4857 (2024).

[28] J. Kim et al., Optimizing cross-domain transfer for universal machine learning interatomic potentials, Nat Commun **17**, 3432 (2026).

[29] M. Neumann, J. Gin, B. Rhodes, S. Bennett, Z. Li, H. Choubisa, A. Hussey, and J. Godwin, *Orb: A Fast, Scalable Neural Network Potential*, arXiv:2410.22570.

[30] B. Rhodes, S. Vandenhaute, V. Šimkus, J. Gin, J. Godwin, T. Duignan, and M. Neumann, *Orb-v3: Atomistic Simulation at Scale*, arXiv:2504.06231.
[31] T. Koker, A. Gangan, M. Kotak, J. Marian, and T. Smidt, *PFT: Phonon Fine-Tuning for Machine Learned Interatomic Potentials*, arXiv:2601.07742.
[32] T. Koker, M. Kotak, and T. Smidt, *Training a Foundation Model for Materials on a Budget*, arXiv:2508.16067.
[33] A. Musaelian, S. Batzner, A. Johansson, L. Sun, C. J. Owen, M. Kornbluth, and B. Kozinsky, Learning local equivariant representations for large-scale atomistic dynamics, Nat Commun **14**, 579 (2023).
[34] A. Hjorth Larsen et al., The atomic simulation environment—a Python library for working with atoms, J. Phys.: Condens. Matter **29**, 273002 (2017).
[35] O. Cohen, J. Riebesell, R. Goodall, A. Kolluru, S. Falletta, J. Krause, J. Colindres, G. Ceder, and A. S. Gangan, TorchSim: an efficient atomistic simulation engine in PyTorch, AI Sci. **1**, 025003 (2025).
[36] J. Riebesell, R. E. A. Goodall, P. Benner, Y. Chiang, B. Deng, G. Ceder, M. Asta, A. A. Lee, A. Jain, and K. A. Persson, *Matbench Discovery -- A Framework to Evaluate Machine Learning Crystal Stability Predictions*, arXiv:2308.14920.
[37] J. P. Perdew, K. Burke, and M. Ernzerhof, Generalized Gradient Approximation Made Simple, Phys. Rev. Lett. **77**, 3865 (1996).
[38] J. W. Furness, A. D. Kaplan, J. Ning, J. P. Perdew, and J. Sun, Accurate and Numerically Efficient $r^2$ SCAN Meta-Generalized Gradient Approximation, J. Phys. Chem. Lett. **11**, 8208 (2020).
[39] J. Schmidt, T. F. T. Cerqueira, A. H. Romero, A. Loew, F. Jäger, H.-C. Wang, S. Botti, and M. A. L. Marques, Improving machine-learning models in materials science through large datasets, Materials Today Physics **48**, 101560 (2024).
[40] L. Barroso-Luque, M. Shuaibi, X. Fu, B. M. Wood, M. Dzamba, M. Gao, A. Rizvi, C. L. Zitnick, and Z. W. Ulissi, *Open Materials 2024 (OMat24) Inorganic Materials Dataset and Models*, arXiv:2410.12771.
[41] H. Kang, T. Lu, Z. Qi, J. Guo, S. Meng, and M. Liu, FastTrack: a fast method to evaluate mass transport in solid leveraging universal machine learning interatomic potential, AI Sci. **1**, 015004 (2025).
[42] G. Cai, Z. Cao, F. Xie, H. Jia, W. Liu, Y. Wang, F. Liu, X. Ren, S. Meng, and M. Liu, Predicting structure-dependent Hubbard U parameters via machine learning, Mater. Futures **3**, 025601 (2024).
[43] Y. Liang, M. Chen, Y. Wang, H. Jia, T. Lu, F. Xie, G. Cai, Z. Wang, S. Meng, and M. Liu, A universal model for accurately predicting the formation energy of inorganic compounds, Sci. China Mater. **66**, 343 (2023).
[44] R. A. Buckingham, The classical equation of state of gaseous helium, neon and argon, Proceedings of the Royal Society of London. Series A. Mathematical and Physical Sciences **168**, 264 (1938).
[45] E. Lee, K.-R. Lee, and B.-J. Lee, An interatomic potential for the Li-Co-O ternary system, Computational Materials Science **142**, 47 (2018).